\documentclass{kluwer}
\usepackage{graphicx}
\begin{document}
\begin{article}
\def\magm{$^{\rm m}\!\!\!.\,$}
\def\EBV{E$_{\rm B-V}$}
\begin{opening}
\title{V4334 Sgr (Sakurai's Nova): The Distance Problem}
\author{S.~Kimeswenger}
\institute{Institut f\"ur Astrophysik, Leopold--Franzens Universit\"at Innsbruck\\ 
Technikerstrasse 25, A-6020 Innsbruck,
             Austria\\}

\begin{abstract}
The central star V4334 Sgr (Sakurai's Nova) of the planetary nebula PN G010.4+04.4 
underwent in 1995-1996 the rare event of a 
very late helium flash.
It represents only one out of two such events during the era of modern astronomy 
(the second event was 
V605 Aql = Nova Aql 1919). All the other prominent
objects of that type  originate from events occurring several thousands of years ago 
(e.g. A30, A78). Thus it is of special interest for stellar evolution theory 
to model the detailed observations obtained during the last four years.
Those models depend essentially on basic stellar parameters like
effective temperature, surface gravity and stellar radius. Most of them
depend strongly on the assumed distance to the object. Some models may
give some constraints on this parameter, but most of them depend on the
assumption as input parameter. 
Hence to determine a relyable distance  is of considerable significance.
This should be obtained through models that give us lower and upper boundaries,
or through means which are independent of models. \\
The detailed review, by using every kind of determination available up to now, 
leads to a  galactic foreground 
extinction of \EBV\,=\,0\magm75$_{\pm0.05}$ and a 
distance of $D\,=\,2.0_{-0.6}^{+1.0}$\,\,kpc.
\end{abstract}
\end{opening}

\sloppy

\section{Introduction}
V4334 Sgr, discovered 1996 February 20 (Nakano et al. \citeyear{Na96}),
was first identified as a nova (Nakano et al. \citeyear{Na96}).
The nature of the object was then identified by Duerbeck \& Benetti (\citeyear{Du96}) 
as the rare event of a 
very late helium flash shell burning ("born-again" scenario; Iben et al. \citeyear{Ib83}; Iben \citeyear{Ib84}).
For a review of this object see Duerbeck et al. (\citeyear{Du00}).
It is only one out of two such events that took place during the era of modern astronomy (the second event was 
V605 Aql = Nova Aql 1919; see the review by Clayton \& de Marco (\citeyear{Cl97}). Older prominent
objects of that type,  where the event occurred several thousands of years ago, are
A30 (Borkowski et al. \citeyear{Bo94}) and A78 (Kimeswenger et al. \citeyear{Ki98}). \\
Thus it is of high interest for stellar evolution theory to be able to model this object.
Generally the stars are described by the three basic parameters effective temperature $T_{eff}$,  
surface gravity $\log(g)$ (or mass) and luminosity $L$. 
While the first value can be determined well and the second reasonably  
well by means of high-resolution spectra,
the third parameter strongly depends  on the 
assumption/measurement of the distance.
"Normally" this is achieved 
by use of photometry and comparison to 
similar sources with known distances. 
V4334 Sgr itself is a prototype object.
We do  not know similar sources.
Obtaining a distance using the luminosity derived
by stellar evolution theory makes all other calculations strongly dependent on the 
chosen model.
Particularly  it does not allow us to verify and improve those theories.
A summary of different methods to derive the distance is given.  
Some of them do not (or only weakly)  depend on
models for this particular object. 
I discuss upper and lower boundaries that one obtains from different theoretical assumptions, and 
attempted to combine them to obtain a reasonable range for the distance to V4334 Sgr. 

\section{The Galactic foreground extinction}
One of the critical parameters for all the model parameters is the galactic foreground extinction.
As we do not know the unreddened colors of the "born-again" core itself, we have to derive
them by other methods. There are basically three methods available.
\vspace{-4pt}{
\begin{enumerate}
\item The diffuse interstellar absorption bands of the foreground on top of the outburst stellar
continuum: This method was applicable in the early phase, where the stellar continuum was 
well defined. The method however suffers from the assumption of average values 
throughout the galaxy. This needs a homogeneous distribution of the gas
 in the galaxy.
K.N. Rao  (Asplund et al. \citeyear{asp97}) suggests a value of \EBV\,$\approx$\,0\magm7 .
\item The hydrogen emission
at various wavelengths and comparison  to theoretical values:
The method is most commonly applied to planetary nebulae, using in our case the old
surrounding nebula, excited by the pre-outburst object.
One is the Balmer decrement.
The theoretical line ratio H$_\alpha$/H$_\beta$ of an (in the UV optically thick 
nebula at an electron temperature above 8\,000 K) 
is known to be $\approx$\,2.85. Thus the measured ration vs. the theoretical one 
gives the foreground extinction.
Duerbeck \& Benetti (\citeyear{Du96}) obtained \EBV\,=\,0\magm56 using a rather poor 
spectrum. Pollacco (\citeyear{Po99}) obtained, with a deeper spectrum, 0\magm71$_{\pm0.9}$. 
The very recent 
determination by Kerber et al. (\citeyear{Ke00}) is based on a very deep VLT spectrum. 
They used, however, a very wide slit.
Thus they have overlaps of the [NII] lines with the H$_\alpha$ line. 
Therefore they do not use only the Balmer decrement, but a photoionization model. 
But this determination is thus not independent from the model (code). They find \EBV\,=\,0\magm79$_{\pm0.6}$.
The uncertainty was calculated by me, taking the data on the Balmer lines from this paper.  
This method 
uses the same emission region, if obtained by a single setup for the spectrograph slit. 
One have to be careful, if the airmass 
of the observation lies above about 1.3. In this case the spectrograph slit has to be 
orientated perpendicular to the horizon.
Otherwise differential refraction causes different regions (clumps) to be observed for 
the H$_\alpha$ and for the H$_\beta$ line.
\item The radio free--free emission ($\lambda$6cm; which gives the number of 
electrons in the 
ionized region) vs. the optical H$_{\beta}$ flux:
This  method
suffers from the fact that we do not know whether we (a) look at the same emission region,
and (b) whether the abundance of heavy elements is normal, since they also  provide also 
electrons for the free--free emission.
Assuming normal conditions we may calculate the expected H$_\beta$ flux $F_\beta$ 
as function of 
the radio flux $F_{\rm 5 GHz}$ ($F_\beta\,\,[{\rm W m^{-2}}]\,=\,3.04\,\,10^{-16}\,F_5\,\,[{\rm mJy}]$,
Ivison et al. \citeyear{Iv91}).
Using now the observed ratio vs. the theoretical ratio we
may obtain the extinction.
Eyres et al. (\citeyear{Ey98b}) obtained $F_5$\,=\,2$_{\pm0.2}$\,mJy and using the
estimate of $F_\beta\,=\,2.08\,\,10^{-17}$\,W\,m$^{-2}$ given by Duerbeck I obtain 
\EBV\,=\,1\magm03$_{\pm 0.04}$
(not 1.13$_{\pm 0.04}$ as given in Eyres et al. \citeyear{Ey98b}).
The errors include the effects on the radio flux only. 
As Duerbeck \& Benetti
did not directly measure the H$_\beta$ flux, 
but used the ratio [OIII]/H$_\beta$ from the 
slit spectrum and the [OIII] image 
assuming a constant [OIII]/H$_\beta$
over the whole nebula, we have to assume a large error.
This is also supported by the findings of Pollacco, who determined a smooth 
H$_\alpha$ distribution and
a rather clumpy [NII] image ([OIII] often goes with [NII]; see A30 and A78 in the HST images).
Using  much deeper images Pollacco (\citeyear{Po99}) derived $F_\beta\,=\,7.0\,\,10^{-17}$\,W\,m$^{-2}$.
But he also used the indirect method described before. 
I obtain \EBV\,=\,0\magm66 using the radio data of Eyres et al. (\citeyear{Ey98b}).
Using the higher radio flux of Eyres (2001),
I obtain an extinction of 0\magm76. The uncertainty in the 
radio and in the H$_\beta$ flux and in the assumptions mentioned above, 
gives an error of up to 0\magm20 . A change of the electron temperature to 12\,000\,K 
(Pollaccco \citeyear{Po99}) leads to a decrease of the value by 0\magm03. 
However, the determination of the electron temperature depends
on the dereddening of the spectrum and thus again on \EBV.
\item Pavlenko et al. (\citeyear{Pav00}) 
test their fitting of the optical spectral energy distribution ($350\,<\,\lambda\,<\,1000$\,nm)
for values of 0\magm54 and 0\magm70.
They conclude, that a value of 0\magm70 is best for this model. As the formation 
of circumstellar dust started before April 1997 and as the derived model parameters $log\,g$ and $T_{eff}$ 
are not independent from the assumed value of \EBV~as input, this determination has to be used carefully.
\item 
Duerbeck et al. (\citeyear{Du00}) argue that the extinction vs. distance diagram 
from Kimeswenger \& Kerber (\citeyear{KiKe98}) indicates nearly no increase of  \EBV\,=\,0\magm8 
from a distance around 2\,\,kpc on. 
Thus they assume this value would be a reasonable choice. 
But this is a purely indirect
argument. 
In particular, since most distance determinations depend even on this value.
\end{enumerate}}
Thus we have to assume 
that the foreground extinction leads to a reddening of  0\magm65\,$\le$\,\EBV\,$\le$\,0\magm85 (conservative). 
I further assume a value of \EBV\,=\,0\magm75.
This good agreement of the results from the optical decrement and from the radio to optical ratio
leads us further to the conclusion that the nebula is optically thick for
the UV photons; otherwise  the theoretical ratio H$_\alpha$/H$_\beta$ would change by 32 percent.
The spread between the different methods is typical for PNe (Gathier et al. \citeyear{Ga86}).

\section{Distances derived from the old planetary nebula}
The old planetary nebula (PN) around Sakurai's star 
may give the best results for the modeling of  the distance.
The main advantages are:
\begin{itemize}
\itemsep=0cm
\item The models for the old PN do not depend on the "born-again" core models.\vspace{-1mm}
\item The PN is not an unusual object and we have thus several hundred objects of that type
for comparison and calibration of the methods used.\vspace{-1mm}
\item As the event occurred just recently, the old nebula should not yet have been affected 
by recombination.\vspace{-1mm}
\end{itemize}

Although there is a huge sample of objects available, the methods for the determination of the
distances are often not very accurate. This mainly originates from predictions and assumptions
for the parameters (mass, electron temperature, filling factor, \dots)  of the ionized gas.
I have sorted the methods in order of decreasing accuracy, although this is somewhat 
a subjective selection. I describe briefly the included assumptions for each method. 
Everyone may do his own selection. 

\medskip
{\bf Extinction - Distance method:}\\
The more simple (and rather inaccurate) version of the method is based on the assumption of a mean interstellar
extinction throughout the galaxy.  
Applying the calibrations of Stasinska et al.  (\citeyear{St93}) I obtain $D\,=\,2.71\,\times\,$\EBV$\,=\,2.0$\,kpc.
This method has to be used carefully.  We have a line of sight, which,
at this distance, resulting in $z = 150\,$pc, leaves the galactic ISM layer. \\
\begin{figure}
\centerline{\includegraphics[width=11cm]{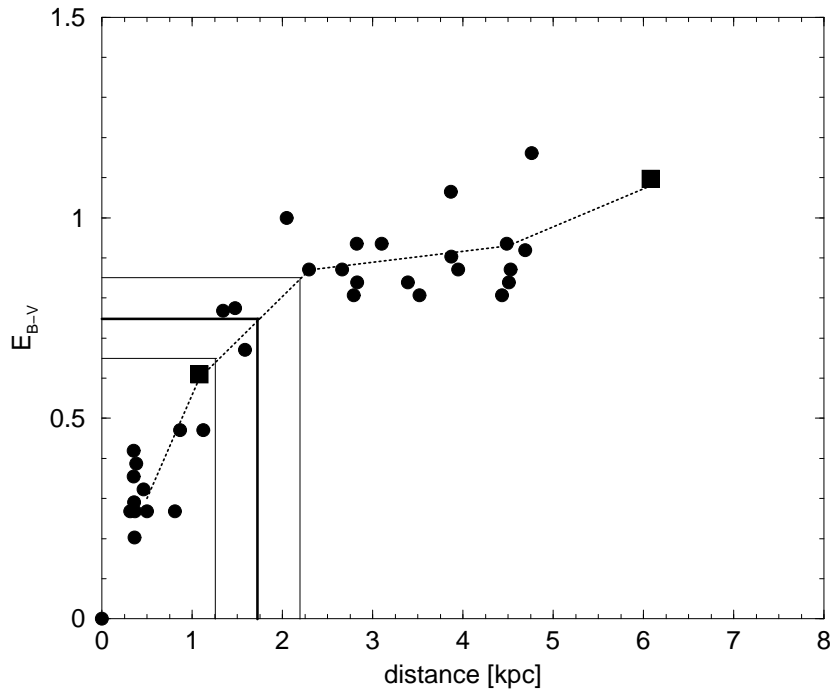}}
\caption{The interstellar reddening \EBV\ as function of the distance  by the photometry of stars in a $r = 2'$ 
field around V4334 Sgr (after Kimeswenger \& Kerber 1998). The two boxes are stars with spectroscopically defined 
type and luminosity class.}
\label{KiKeDist}
\end{figure}\\
The more reliable method is the one, introduced by Lutz (\citeyear{Lu73}), Acker (\citeyear{Ac78}) 
and Lucke (\citeyear{Lu78}), which is using the individually derived
extinction distance diagram of stars in the field around the object.
But their determinations suffer from the fact, that they were using only literature data of photometrically
known stars. Thus their typical field radius was up to two degrees. At such scales, 
the ISM is too clumpy. Only for a small number of PNe the deep photometry of the stars in a small field
(few arcminutes) was used to derive a local extinction profile (e.g. Gathier et al. \citeyear{Ga86}).
The main assumptions are, that the 
interstellar extinction is homogeneous within the field, 
and that the extinction derived in the nebula originates from the ISM foreground and not from
internal extinction. A radius $r\,=\,2'$ gives only a beam of 1.2\,pc at
a distance of 2\,kpc (a few times the PN size). Therefore the first assumption seems to be reasonable.
The nebula is rather thin ($n_e\,\le\,300$). Even a low gas to dust ratio 
will not result in a strong contribution from the PN internally. \\
This leads, using the new values for \EBV\ and the photometry of Kimeswenger \& Kerber
(\citeyear{Ki98}) to a distance of\\
\phantom{XXXXX}
\begin{tabular}{l c l l}
$D_{\rm KK}$ & = & 1.7$_{-0.5}^{+1.0}$ kpc &\\
\end{tabular}\\ 
This is the only method, which works without assumptions on the physics of the nebula 
(mass, density, ionization rate, \dots)  and without assumptions on the homogenity of galactic 
parameters (rotation, overall ISM distribution,  \dots).
It is only influenced by the
accuracy of the measurements (photometry of the stars around and the determination of the
\EBV\ of the nebula).

\medskip
{\bf Galactic kinematics:}\\
The majority of the Population II objects follow the galactic rotation. Thus this might be used to 
derive the distance by using the radial velocity of V4334 Sgr. But this determination 
includes two assumptions: {\begin{itemize}\vspace{-4pt}
\item The progenitor star of the object was a  disk population star.
\vspace{-5pt}
\item  The object does not have a peculiar orbit.
\end{itemize} \vspace{-4pt}}
Both assumptions, although being statistically  correct for samples of objects, 
are dangerous
for individual stars. We know about disk stars which do have large velocities, 
with respect to the surrounding field,
of up to 150\,km\,s$^{-1}$.
There are also bulge objects known in the solar 
neighborhood. 
 Zijlstra et al. (\citeyear{zi97}) studied the
radial velocity distribution of PNe towards the galactic bulge. They conclude, that PNe
with distances $<$~4~kpc
follow the galactic rotation (velocity dispersion $\sigma$\,=\,46\,km\,s$^{-1}$), 
while those at higher distances have deviations of up
to 250\,km\,s$^{-1}$  from the radial velocity field 
($\sigma$\,=\,114\,km\,s$^{-1}$). However their sample of PNe contains
only 7 objects below 4 kpc. The object M1-20, which is the nearest to V4334 Sgr in this 
sample,
also has a strong deviation from the local galactic rotation field.  
Currently 12 PNe are catalogued in the vicinity (r = 2$^{\rm o}\!\!.\,$5) of V4334 Sgr. 
Six have reliable distance determinations (3 of them very reliable).
Most of them deviate strongly from the galactic rotation (standard deviation 64\,km\,s$^{-1}$;
using only objects with D $<$ 4kpc: standard deviation 73\,km\,s$^{-1}$).
The object with the best distance determination (NGC 6439; 1.1 kpc) has a deviation of 114\,km\,s$^{-1}$
from the galactic rotation.
At an assumed distance of 3\,kpc the deviation of V4334 Sgr from the galactic rotation field is 70\,km\,s$^{-1}$.
Thus this method can not be used for a distance determination for an individual object 
in this direction of the galaxy.

\medskip
{\bf The statistical distances using the 6\,cm radio flux:}\\
The simple versions of those methods are based on the assumption, that the nebulae 
are ionized regions with a constant abundance and a constant ionization rate for all 
elements throughout the nebula.
Then the radio free-free flux is thus connected to the mass, the distance of the 
nebula and the electron temperature. Assuming a mass and an electron temperature 
we may derive the distance
(Milne \& Aller \citeyear{MiAl75}). The electron temperature
has almost no affect in the equation (power of -0.1) and thus may be neglected.
The typical mass of old PN shells is 0.1\,M$_\odot$. The mass enters with a
 power of $^2/_5$. Having a range of $0.03\,<\,M_{\rm PN}\,<\,0.3$ the distance $D$ 
derived in such way
may deviate by about 40 percent. This method assumes a filling factor of 0.6 .
This might be too high for such a low surface brightness object. Assuming a filling factor of 0.2
reduces the distance by about 20 percent.
The more sophisticated methods are based on Daub (\citeyear{Dau82}), where a nebula is first
a ionization bound compact object (not the whole shell is already ionized) and later the
nebula is density bounded and thus has a constant ionized mass (the whole nebula is ionized).
Different variations of this method were developed. They are then either calibrated on 
nearby PNe with known distances (Cahn et al. \citeyear{CSK}, hereafter CSK) or by the so called bulge 
sample, assuming a constant distance for those objects (van der Steene \& Zijlstra \citeyear{VSZ}, hereafter VSZ).
The probably most sophisticated version is that given by Schneider \& Buckley (\citeyear{SB96}). 
They use theoretical models of the radio surface brightness and 
support it by numerical simulations on samples of PNe including different physical parameters like
masses, filling factors, luminosities, \dots. They calibrate also
with the bulge sample, but use the bulge as an extended distribution. They also 
describe, why VSZ and CSK have problems with extended low surface density objects. 
They found that the previous methods
{\bf systematically overestimate} distances at such low surface brightness objects like the 
PN around V4334 Sgr. \\
Using the radio flux of 2.7 mJy (Eyres 2001) the four commonly used methods give: \\
\phantom{XXXXX}
\begin{tabular}{l c l l}
$D_{\rm MA}$ & = & 3.0 kpc &(Milne \& Aller \citeyear{MiAl75})\\
$D_{\rm CKS}$ & = & 3.6 kpc & (Cahn et al. \citeyear{CSK})\\
$D_{\rm VSZ}$ & = & 6.9 kpc & (van der Steene \& Zijlstra \citeyear{VSZ})\\
$D_{\rm SB}$ & = & 3.5 kpc & (Schneider \& Buckley \citeyear{SB96})\\
\end{tabular}\\
As those methods tend to overestimate the distance and as we
have to take into account that the filling factor and the mean mass
assumed for those calculations is
too high, we may conclude that the distance derived by means of those methods
is 2.0 to 3.5 kpc. The uncertainty of this method for individual objects is rather high 
(see comparison in Fig.3 in Jacoby et al. \citeyear{Ja98}).

\medskip
{\bf The statistical distances using optical $H_\beta$ fluxes:}\\
Those methods are including very similar assumptions than the radio flux methods.
They do not suffer from the assumption on the (unknown) He$^{++}$ contribution.
On the other hand they depend strongly on the ISM reddening \EBV.\\
The original method by Shklovskii (\citeyear{Sh56}) was based on the surface brightness
at POSS plates. The direct application to it is the method of O'Dell (\citeyear{OD62}).
Maciel \& Pottasch (\citeyear{MaPo80}) find in their sample a good correlation
of the mean H$_\beta$ surface brightness vs. linear radius of the object. Pottasch (\citeyear{Po84})
gives a slightly different application using a mean mass. I also used his method applying 
the whole mass range mentioned above. Assuming the (conservative) range 0\magm70$\,\le\,$\EBV\,$\le\,$0\magm85
and an error of 50 (!) percent for the H$_\beta$ flux of Pollacco (\citeyear{Po99}) I derive \\
\phantom{XXXXX}
\begin{tabular}{l c l l}
$D_{\rm MP}$ & = & 3.7$_{-1.2}^{+0.8}$ kpc &(Maciel \& Pottasch \citeyear{MaPo80})\\
$D_{\rm OD}$ & = & 2.9$_{-0.4}^{+1.2}$ kpc & (Shklovskii \citeyear{Sh56}; O'Dell \citeyear{OD62})\\
$D_{\rm POa}$ & = & 2.2$_{-0.2}^{+0.3}$ kpc & (Pottasch \citeyear{Po84}; constant mass)\\
$D_{\rm POb}$ & = & 2.3$_{-0.7}^{+0.4}$ kpc & (Pottasch \citeyear{Po84}; mass range)\\
\end{tabular}\\
The newer surface brightness based methods give lower distances at higher \EBV. This is
exactly contrary to the extinction distance  (see Fig. 1). Thus the combination of
those methods gives us some
supplementary information for the value of \EBV. These methods are based/calibrated using mean masses
and filling factors of the samples of PNe available at that time. Thus the extremely evolved
PN of V4334 Sgr might have a lower mass and filling factor and thus a slightly lower distance.
\begin{figure}
\centerline{\includegraphics[width=10.00cm]{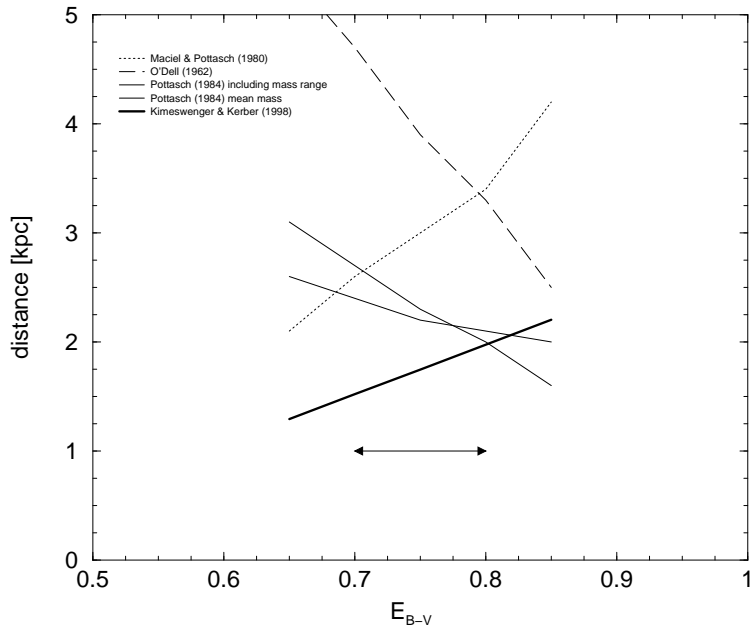}}
\caption{The distance as function of the reddening \EBV using the H$_\beta$ flux.}
\end{figure}

\medskip
{\bf The "photometry" of the central star of the PN - \\\indent The progenitor of V4334 Sgr:}\\
The modeling of the PN gives us an idea about the luminosity and the effective temperature of the
progenitor star. Pollacco (\citeyear{Po99}) gives 80\,000\,$\le\,T_{eff}\,\le\,$200\,000. This results in an inaccuracy
(assuming a constant luminosity) of 2\magm9 for the V magnitude. Kerber et al. (\citeyear{Ke00})
give 100\,000$_{\pm 7\,000}$\,K. This would reduce the uncertainty to 0\magm4 (at constant luminosity).
More severe is the estimate on the luminosity. Pollacco is giving $L\,=\,2\,000\,$L$_\odot$ but including a range 
from 100 to 40\,000\,L$_\odot$ ($\Delta\,m_V\,\approx\,3$\magm3 at given distance and temperature). 
Kerber et al. are giving a range resulting
in $\Delta\,m_V\,\approx\,2$\magm5. Thus the absolute magnitude from the model is not better than about 3$^{\rm m}$.
Do we compare that with the possible detection of the progenitor with J\,=\,21$^{\rm m}$ we obtain a 
distance of \\
\phantom{XXXXX}
\begin{tabular}{l c l l}
$D_{\rm CSPN}$ & = & 2.5$_{-1.5}^{+2.0}$ kpc &\\
\end{tabular}\\
The uncertainties are very high, because the progenitor is hardly detected, and since 
the luminosity and the effective
temperature is very uncertain.

\section{Distances derived from models of V4334 Sgr}

The "born-again" core V4334 Sgr allows us also to derive distances by using models.
On the one hand there are models of the final helium flash. They imply, that the 
star should reach a luminosity of 5\,000 to 20\,000\,L$_\odot$ (depending on the assumed mass of the core).
A problem is the "speed" of the evolution. We also do not know exactly, when the
star should be "back" at those luminosity. Assuming that it happened not before end of 1997
I obtain (for $L\,=\,8\,000\,$L$_\odot$) a distance of 2.5 to 3.0 kpc (Kimeswenger \& Koller \citeyear{Ki00this}).\\
Another access is the modeling of the stellar photosphere in 1996. Those models give
us an effective temperature and the surface gravity. Assuming a mass (it depends only weakly on the mass)
we may obtain a luminosity and thus a distance. Up to now, those models seem to be not accurate
enough, to really obtain the values properly. The work of Asplund et al. (\citeyear{asp97}) gives a decrease
of the stellar luminosity by a factor of 4 during 1996. The photometry gives us an increase of about a factor of 2.
Thus the error on those models, directly going to the distance, are about one order of magnitude.
This field will evolve rapidly  in the next future.

\section{Trigonometric methods - an outlook}

Direct trigonometric methods are providing the mostly model independent distances.
Up to now it was not possible to obtain appropriate data, but 
there are some (weak) prospects for the future.

\medskip
{\bf Annual Parallax:}~\\
As we expect this object to be at a distance significantly above 1\,kpc, classical work with annual parallaxes do not work
at the moment. So we can do a forecast only.
Although the DIVA mission (start 2003) has a extremely red sensitive detector allowing a red object like 
Sakurai to be measured down to a visual magnitude of about 19\magm0 the current fading hides this unique
prototype object for this satellite. If we assume recovery like it is known for FG Sge and the R CrB stars
we do have some hope. If V4334 Sgr follows more the track of V506 Aql, it will not brighten anymore during my lifetime.
GAIA (start 2010 ?), which achieves 0.1 mas and completeness at V=20 for blue stars and more for red stars,
will measure it, if it does not fade further more.

\medskip{\bf Proper Motion}~\\
Sakurai's nova has a high radial velocity. If the distance of the object is below e.g. 5kpc we should have radial velocities
due to the galactic rotation $\approx$30\,km\,s$^{-1}$.
The measured velocity is about 90\,km\,s$^{-1}$ higher. Assume it is a random
peculiar motion the probability 
that it has half of this value or above perpendicular of the 
line of sight is high. Thus it would have a proper motion of a few tenths of arcseconds per century.
This may give us an upper limit for the distance.

\medskip{\bf Angular Size}~\\
As discussed above the expansion of the newly formed shell should be at 200--500\,km\,s$^{-1}$. The hot dust is concentrated
in the core only (Kimeswenger \& Koller \citeyear{Ki00this}). Thus we either need speckle interferometry at longer wavelengths 
($>$ 10$\mu$m)
or radio interferometry of the gas in the core. This should be able to resolve the region, assuming the expansion 
started end of 1996 or early in 1997 and that the object is within 3 kpc.

\section{Conclusions}
The distance determinations still allow a wide variety of values. Trigonometric methods are not available in the 
nearby future, but upper
limits from proper motion might be possible.
\begin{figure}
\centerline{\includegraphics[width=10.00cm]{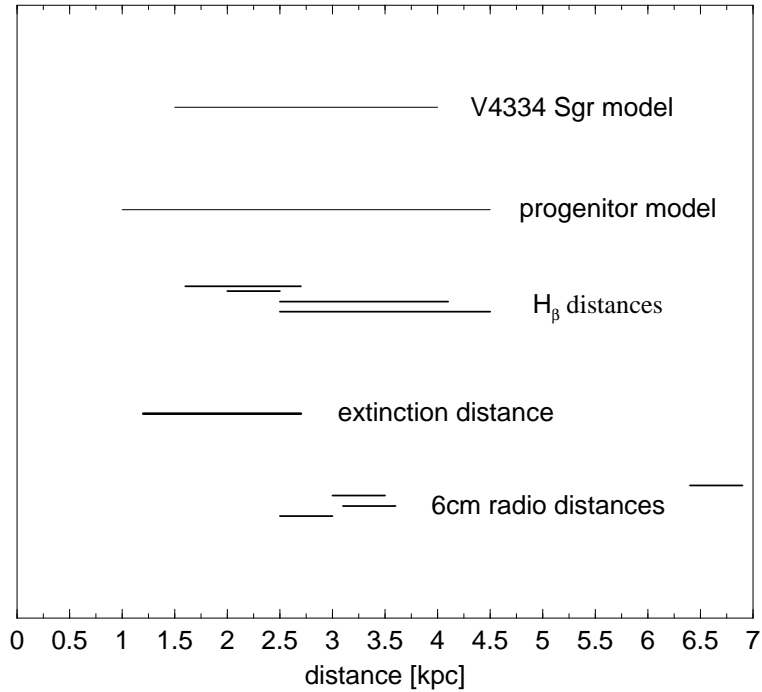}}
\caption{ The different distance determinations for V4334 Sgr grouped according to the method.}
\end{figure}
I conclude that the interstellar foreground reddening is most likely between 0\magm7 and
0\magm8. The distance thus remains somewhat uncertain, but we have reasons to assume that it lies between 1.5 and 3 kpc.
For the moment I suggest (after extensive discussions with collegues)
to use always two "canonical" values for grids of theoretical models - 2.0 and 5.0 kpc. \\
A distance of 2\,kpc leads to a bolometric luminosity of about 4\,500\,L$_\odot$ for V4334 Sgr at end of 1997 and
a nebular mass (after Milne \& Aller \citeyear{MiAl75}) of 0.05$_{\pm 0.02}\,$M$_\odot$. The uncertainty reflects 
mainly the range of the filling factor. The resulting linear size is 0.4\,pc.
2.5\,kpc (3.0\,kpc) lead to 7\,000 (10\,000) L$_\odot$ and to a PN mass of 0.08$_{\pm 0.03}\,$ (0.12$_{\pm 0.03}\,$) M$_\odot$.
Those values are very reasonable for this type planetary nebula. Nevertheless keep in mind that a low luminosity slows down the born-again
evolution drastically.
This slow evolution in the models seems to be the main problem in our understanding of this event.

\acknowledgements
This  project  was  supported by the Austrian FWF  project P11675-AST.
In particular I would like to thanks
the LOC of the meeting in Keele for their invitation to this talk.

\end{article}
\end{document}